# DATA EFFICIENT VOICE CLONING FOR NEURAL SINGING SYNTHESIS

*Merlijn Blaauw\*, Jordi Bonada\*, and Ryunosuke Daido†*

\*Music Technology Group, Universitat Pompeu Fabra, Barcelona, Spain
†Sound Processing Group, Yamaha Corporation, Hamamatsu, Japan
merlijn.blaauw@upf.edu, jordi.bonada@upf.edu, ryunosuke.daido@music.yamaha.comabstract**ABSTRACT**

There are many use cases in singing synthesis where creating voices from small amounts of data is desirable. In text-to-speech there have been several promising results that apply voice cloning techniques to modern deep learning based models. In this work, we adapt one such technique to the case of singing synthesis. By leveraging data from many speakers to first create a multispeaker model, small amounts of target data can then efficiently adapt the model to new unseen voices. We evaluate the system using listening tests across a number of different use cases, languages and kinds of data.

*Index Terms*— Singing synthesis, voice cloning, speaker embedding, speaker adaptation, multispeaker model## 1. INTRODUCTION

In singing synthesis, the goal is often to reproduce a specific target singer's voice, given an input such as a musical score with lyrics. In this context, the ability to create a new voice from a small amount of recordings, e.g., 2 min, is desirable for many use cases. There may, for instance, be cases where no new material can be obtained because the target singer is already deceased. Other cases might require modeling a large number of vocal timbres where recording a full dataset for each voice would be impractical, such as for large choir synthesis.

Voice cloning, also known as voice fitting or speaker adaptation, is a technique that leverages data from many speakers[1] combined with a small amount of data from the target speaker to allow creating a voice model that outperforms a model trained on just the adaptation target data from scratch. The recent wave of text-to-speech (TTS) systems based on modern deep learning techniques not only achieve state of the art results [e.g., 1, 2], they also make voice cloning relatively straight forward.

In this work, we apply voice cloning techniques to singing synthesis using a modern singing synthesizer based on an autoregressive neural network architecture [3]. Here, we focus on cloning timbrical aspects of the target voice only, not considering expressive aspects related to the interpretation of the given musical score. Instead, we consider pitch and phonetic timings control inputs given by some external source, such as a recording [e.g., 4], or another model [e.g., 5, 6]. While simplifying the problem, we argue such an approach is still useful for many practical applications, and should provide greater data efficiency.

The main contributions of this paper are (i) adaptation of voice cloning techniques used in TTS to a singing synthesizer, (ii) evaluation of several practical aspects of singing voice cloning, such as what kind of data to use, across datasets in English, Japanese and Spanish/Catalan.

## 2. METHODOLOGY

We first briefly review the relevant parts of our baseline singing synthesizer, and then discuss the extension to multispeaker modeling and speaker adaptation.

### 2.1. Preliminaries

Our baseline singing synthesizer described in [3] maps time-aligned control features, $\mathbf{c} = [\mathbf{c}_1, \mathbf{c}_2, \ldots, \mathbf{c}_T]$, to acoustic vocoder features, $\mathbf{x} = [\mathbf{x}_1, \mathbf{x}_2, \ldots, \mathbf{x}_T]$, using an autoregressive model,

$$p(\mathbf{x}|\mathbf{c}) = \prod_{t=1}^{T} p(\mathbf{x}_t|\mathbf{x}_{<t}, \mathbf{c}_t; \theta). \quad (1)$$

Here, $p(\mathbf{x}_t|\mathbf{x}_{<t}, \mathbf{c}_t; \theta)$ is an isotropic output distribution whose parameters are predicted using a slightly modified WaveNet architecture parameterized by $\theta$.

### 2.2. Multispeaker model

In order to leverage data from many speakers, we first extend this basic model to a multispeaker model. Such a model is a single network able to model all voices in the training data simultaneously, by conditioning it on an additional low-dimensional vector representing the different speaker identities. We follow [7] and use a learned speaker embedding which depends on acoustic properties of each speaker. We thus extend the model in eq. (1) to become,

$$p(\mathbf{x}|\mathbf{c}, i) = \prod_{t=1}^{T} p(\mathbf{x}_t|\mathbf{x}_{<t}, \mathbf{c}_t, i; \theta, \mathbf{s}), \quad (2)$$

---

[1]Throughout this document, the term "speaker" is used regardless of whether the subject is speaking or singing.

where $\mathbf{s} = [\mathbf{s}_1, \mathbf{s}_2, \ldots, \mathbf{s}_N]$ is a speaker embedding table consisting of an $M$-dimensional speaker vector for each of the $N$ speakers. These vectors are initialized using a uniform random in the range $[-0.1, 0.1]$, and then learned jointly with the other model parameters, $\theta$, during training. The model is additionally conditioned on a speaker index, $i$, used to select the corresponding speaker embedding vector, $\mathbf{s}_i$, from the speaker embedding table. This vector is then used to actually condition the generator, in our case by simply appending the speaker embedding vector to the control vector for each timestep,

$$\hat{\mathbf{c}}_t = [\mathbf{c}_t; \mathbf{s}_i]. \qquad (3)$$

The underlying idea here is that if we train such a model on sufficiently many speakers, it should be possible to generalize to new speakers by simply finding the corresponding speaker embeddings, but keeping all the other weights in the model fixed.

### 2.3. Speaker adaptation

In order to adapt a multispeaker model to a new unseen speaker, a randomly initialized vector is added to the speaker embedding table, and the training of the model is continued on data of the new speaker, using the corresponding speaker index and generative loss as usual. In this case, the multispeaker weights, $\theta$, are kept fixed and only the new speaker embedding, $\mathbf{s}_{i=N+1}$, is updated. However, as previously reported in e.g., [8, 9], the generalization capabilities of a multispeaker model tend to be limited in practice, and better results in terms of synthesis quality and speaker similarity are obtained using so-called *finetuning* of all the model weights.

When finetuning the whole model, both the speaker embedding, $\mathbf{s}_i$, and the multispeaker model weights, $\theta$, are updated using the generative loss computed over data of the new speaker. As the goal of voice cloning is to use very few material of the target voice, this data tends to have a very small size. Thus, unlike learning only the new speaker embedding, finetuning the whole model tends to overfit. A simple solution is to use early stopping and only finetune the whole model for a small number of updates. In contrast, learning a new speaker embedding tends to converge after many more updates. Intuitively, it therefore makes sense to first optimize the speaker embedding for many updates and then do a finetuning of the multispeaker model weights for few updates. However, we empirically found that this approach does not clearly benefit the results compared to jointly learning all parameters for few updates, and thus we tend to use this latter approach as it is much faster.

### 2.4. Singing voice specifics

Compared to speech, expressive singing tends to span a wider range of pitches and timbres. As our proposed system uses a vocoder [10] which separates pitch and timbre to a great extent, generalizing to a wide range of pitches can be done efficiently. However, the high degree of variability in terms of timbre should be considered, especially since datasets in singing synthesis tend to have less data for the single speaker case or less speakers for the multispeaker case, compared to datasets in TTS.

In order to keep the intraspeaker timbre variability manageable, we propose to use what we call *pseudo singing* for the multispeaker model. Pseudo singing are phrases that are sung with approximately constant pitch, cadence and dynamics, forcing a more clear and coherent pronunciation. Using this kind of data thus provides a convenient way of ensuring a more homogeneous base voice when data is limited. The recordings used for the target voice to clone can be either natural or pseudo singing.

### 2.5. WaveNet vocoder

The use of a traditional heuristics-based vocoder has some benefits in the case of singing synthesis, but at the same time places an upper bound on the obtainable sound quality. Especially when the used generative model introduces some form of smoothing, the synthesized waveform can often sound overly "buzzy". One way to mitigate these problems is to use a neural vocoder, e.g., a WaveNet model that predicts waveform from vocoder features [11, 12]. While the amount of data required to train such a WaveNet vocoder may seem contrary to the goals of voice cloning, we have obtained promising results by combining data from different speakers in order to learn what is sometimes called a *universal* mapping of vocoder features to waveform [13, 14]. As training data in this case only consists of audio, without requiring text, speaker identities or other annotations, the burden of collecting such a dataset is notably reduced.

## 3. EXPERIMENTS

### 3.1. Datasets

The datasets used in this work, consisting of audio with aligned phonetic transcription and speaker identities, are listed in table 1. The phonetic transcription and segmentation was done automatically for the pseudo singing datasets, but some manual correction was required in the case of natural singing. Except for `NUS-48E` [15], all datasets are proprietary.

### 3.2. Experimental conditions

The hyperparameters used for the experiments generally follow [3], but were simplified slightly and the number of channels was increased to increase capacity for the multispeaker model. In particular, we use an initial $1 \times 1$ convolution, followed by 5 causal $2 \times 1$ convolutional layers with gated tanh nonlinearity, dilation factors $\{1, 2, 4, 8, 16\}$, and non-parameterized residual connections. These layers use 384 or 256 channels, and 256 or 128 skip channels for multispeaker and single speaker models respectively. The output stack is comprised of a fully connected ReLU layer, a fully connected output layer and a simple L1 loss, rather than a mixture density negative log likelihood loss. We used a 16-dimensional speaker embedding for all experiments. The multispeaker models are trained for around 800 k iterations, with an initial learning rate of $3 \times 10^{-4}$. Adaptation finetuning is done for 4 k additional iterations, applying Polyak averaging [16]. The baseline single speaker models are trained for 30–90 k iterations with a learning rate of $5 \times 10^{-4}$.

**Table 1.** The different datasets used in the experiments and their properties.

| Tag | Speakers | Kind | Language | Gender | Style | Size[1] | Pitches | Avg. dur.[2] |
|---|---|---|---|---|---|---|---|---|
| `JP-MULTI-P` | 35 | Pseudo | Japanese | 8M/27F | Mixed | 80 | 3 | 0:14:48 |
| `EN-MULTI-P` | 13 | Pseudo | English, mixed native | 8M/5F | Mixed | 524 | 12×1,1×2 | 0:32:28 |
| `NUS-48E` [15] | 12 | Natural | English, non-native | 6M/6F | Pop | 4 | - | 0:10:35 |
| `ES-MULTI-A,B-P` | 2×8[3] | Pseudo | Spanish, Catalan | 2×4M/4F | Choir | 219 | 3 | 0:51:47 |
| `JP-TAR-K-P` | 1 | Pseudo | Japanese | Female | Pop | 9 | 3 | 0:01:49 |
| `EN-TAR-PS-N` | 1 | Natural | English | Male | Pop | 2 | - | 0:03:31 |
| `EN-TAR-AM-N` | 1 | Natural | English | Female | Pop | 6 (excerpts) | - | 0:02:14 |
| `EN-TAR-AM-P` | 1 | Pseudo | English | Female | Pop | 10 | 3 | 0:02:23 |
| `ES-TAR-*-P` | 2×4×1[3] | Pseudo | Spanish, Catalan | 2×2M/2F | Choir | 10 | 3 | 0:02:22 |
| `JP-FULL-K-P` | 1 | Pseudo | Japanese | Female | Pop | 80 | 3 | 0:14:35 |
| `EN-FULL-PS-N` | 1 | Natural | English | Male | Pop | 39 | - | 1:29:01 |

[1] Size in number of utterances per speaker per pitch for pseudo singing datasets, or number of songs per speaker for natural singing datasets.
[2] Average duration per speaker (combining all pitches) for multispeaker datasets, or total duration of single speaker datasets.
[3] In the experiments, from total of 12 speakers, we create two 8 speaker multispeaker models with a different set of 4 speakers held out. Each of these models is then adapted to its 4 held out speakers, for a total of 8 targets.

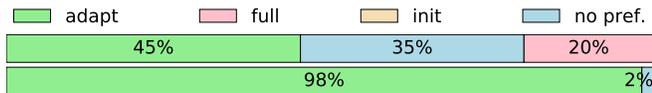

**Fig. 1.** Results of test comparing adapted voice ("adapt", 1 min 49) to a voice trained on full dataset ("full", 14 min 48) and a voice trained on adaptation data from random initialization ("init", 0% preference) respectively. In this case with Japanese pseudo singing multispeaker and target data.

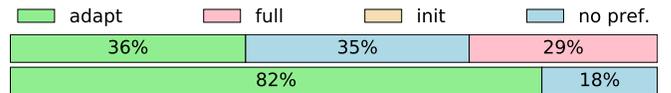

**Fig. 2.** Results of test comparing adapted voice ("adapt", 3 min 31) to a voice trained on full dataset ("full", 1 h 29) and a voice trained on adaptation data from random initialization ("init", 0% preference) respectively. In this case with English pseudo singing multispeaker data and natural singing target data.

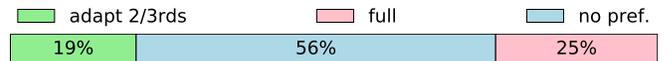

**Fig. 3.** Results of test comparing a synthetic choir consisting of 8 adapted voices and 4 full dataset voices ("adapt 2/3rds", 8 × 2 min 22, 4 × 51 min 47) to a choir trained on full datasets ("full"). In this case with Spanish/Catalan pseudo singing multispeaker and target data.

### 3.3. Listening tests and results

We conducted a series of listening tests to evaluate the proposed system perceptually, in lieu of reliable quantitative metrics. The listening tests consisted of simple AB preference tests, where participants were asked to select the preferred stimulus ("A", "B" or "no preference"), considering a reference recording of the target singer. In all tests this reference recording is also used to control pitch and phonetic timings. In total there were 19 participants, each rating 20 pairs of acapella stimuli, divided over 5 tests.

The first test tries to measure the effectiveness of voice cloning, in the case of a Japanese pseudo singing multispeaker model, `JP-MULTI-P` (see table 1), and a pseudo singing adaptation target, `JP-TAR-K-P`. The test compares the adapted model to a model trained on the full dataset, `JP-FULL-K-P`, and the adapted model to a model trained from random initialization on the same small adaptation dataset. In fig. 1, we can see that the adapted model actually outperforms the model trained on the full dataset. One explanation could be that the relatively large amount of data used to train the underlying multispeaker model improves overall sound quality, while both models capture the target speaker identity similarly well. As expected, the adapted model is consistently preferred over the model trained on a small amount of data from random initialization.

The second test is similar to the first test, but for the case of an English pseudo singing multispeaker model, `EN-MULTI-P`, a natural singing adaptation target, `EN-TAR-PS-N`, and a full dataset of natural singing, `EN-FULL-PS-N`. In fig. 2, we see that the adapted model and the full dataset model perform similarly. This shows that in this case cloning is as effective as training on a larger dataset. Again, the adapted model is consistently preferred over the model trained on a small amount of data from random initialization.

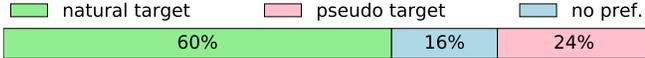

**Fig. 4.** Results of a test comparing adaption of a pseudo singing multispeaker model to natural singing (2 min 14) and pseudo singing (2 min 23) respectively. In this case with English data.

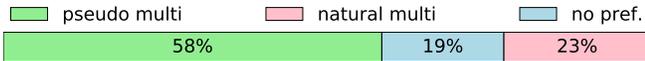

**Fig. 5.** Results of a test comparing adaption of a multispeaker model from pseudo singing (13 × 32 min 28) or from natural singing (12 × 10 min 35) to natural singing. In this case with English data.

The third test considers the use case of choir singing. In particular, using cloning to create larger choirs, e.g., 8 full voices and 24 cloned voices adapted from the former set of full voices. As the available dataset consisted of just 12 voices, 3 per soprano, alto, tenor and bass part, we performed a scaled-down experiment, generating a choir of 4 full voices and 2×4 cloned voices, adapted from two different 8 voice multispeaker models. First, a multispeaker model is trained on 8 voices, `ES-MULTI-A,B-P`, holding out one voice per part. Then, adapted models are created for each of the held out voices, `ES-TAR-*-P`, and the whole process is repeated for a different set of held out voices. Finally, we generate the choir synthesis by mixing 4 full voices and 8 cloned voices. We compare this result with a result of a single multispeaker model trained on the full dataset of all 12 voices. In fig. 3, we can see that both models perform on-par, showing that in this case using cloned voices does not significantly degrade the result.

The fourth test compares using a natural singing adaptation target, `EN-TAR-AM-N`, to using a pseudo singing adaptation target, `EN-TAR-AM-P`, in the case of English. Both use a multispeaker model trained on pseudo singing, `EN-MULTI-P`. In fig. 4, we see that a natural singing target is preferred. One explanation for this is that pseudo singing tends to be a little overpronounced compared to natural singing, thus producing a timbre a little farther from the reference stimulus.

The fifth test compares using a multispeaker model trained on pseudo singing, `EN-MULTI-P`, to a multispeaker model trained on natural singing, `NUS-48E`, in the case of English. Both are then adapted to natural singing. In fig. 5, we see that a multispeaker model trained on pseudo singing is preferred. One possible explanation for this could be that the more coherent pseudo singing provides a more homogeneous base voice. However, as the two multispeaker datasets have different sizes and consist of different speakers, it is difficult to draw any definitive conclusions from this single result.

A number of sound examples are available online[2].

## 4. RELATION TO PRIOR WORK

There have been several recent works on voice cloning. Following [8], the two main approaches are *speaker adaptation* and *speaker encoding*. Speaker adaptation, which is used in this work, optimizes a speaker embedding for a new speaker within a multispeaker model using gradient descent. Speaker encoding uses a secondary network to predict a new speaker embedding from acoustic features. The latter approach has the advantage that voice cloning only requires a single forward pass rather than a costly iterative optimization, and that no transcription of the acoustic adaptation material is needed. However, both approaches have been shown to benefit significantly from a finetuning of all model weights [8, 9], effectively negating these benefits from the latter method.

Arik *et al.* [8] compare speaker adaptation and encoding approaches for the Deep Voice 3 model, which is a convolutional sequence-to-sequence model that predicts mel-spectrogram features combined with a Griffin-Lim vocoder [17]. Jia *et al.* [13] propose an encoding approach in context of the Tacotron 2 synthesizer, using speaker vectors obtained using a network trained for speaker verification using a discriminative loss rather than a generative loss. In this case the waveform is generated from predicted mel-spectrogram features using a multispeaker WaveNet vocoder. Taigman *et al.* [18] and Nachmani *et al.* [19] discuss speaker adaptation and encoding respectively, in context of the novel shifting buffer VoiceLoop model, with the latter using a jointly optimized generator and encoder network using a combination of generative, contrastive and cycle constancy losses. Here, like in our work, the model predicts vocoder features. Similarly, Chen *et al.* [9] use speaker adaptation and speaker encoding using *d-vectors* (as in [13]), but adapt these techniques to the WaveNet model, notably using waveform directly rather than mel-spectrogram or vocoder features. There are several small architectural differences between these works, such as how conditioning on speaker embedding is implemented.

The approach proposed in this paper has similarities to many of the above works. However, we apply these techniques to the model proposed in [3], and evaluate the model in a context relevant to singing rather than speech. Related prior works in singing synthesis have been limited to speaker adaptive training of hidden Markov models [20] to increase data efficiency.

## 5. CONCLUSIONS

The presented results for voice cloning applied to singing voice are encouraging. From small amounts of data, the target speaker identity can be convincingly reproduced, while maintaining a sound quality comparable to non-cloned voices. One convenient approach is to combine a multispeaker model trained on pseudo singing, with a natural singing adaptation target. While the proposed system, focused on cloning timbre, has many practical applications, we ultimately want to clone both timbrical and expressive aspects of the target voice. In this case, going away from a traditional TTS pipeline and towards an end-to-end system would be a promising direction.

---

[2]https://mtg.github.io/singing-synthesis-demos/voice-cloning/

## 6. REFERENCES

[1] A. van den Oord, S. Dieleman, H. Zen, K. Simonyan, O. Vinyals, A. Graves, N. Kalchbrenner, A. W. Senior, and K. Kavukcuoglu. (2016). WaveNet: A generative model for raw audio. version 2. arXiv: 1609.03499 [cs.SD].

[2] J. Shen, R. Pang, R. J. Weiss, M. Schuster, N. Jaitly, Z. Yang, Z. Chen, Y. Zhang, Y. Wang, R. Skerry-Ryan, R. A. Saurous, Y. Agiomygiannakis, and Y. Wu, "Natural tts synthesis by conditioning WaveNet on mel spectrogram predictions," in *Proceedings of the 43rd IEEE International Conference on Acoustics, Speech, and Signal Processing (ICASSP)*, Calgary, Alberta, Canada, 2018, pp. 4779–4783.

[3] M. Blaauw and J. Bonada, "A neural parametric singing synthesizer modeling timbre and expression from natural songs," *Applied Sciences*, vol. 7, no. 12, 1313, V. Välimäki, Ed., Dec. 2017, *Special Issue on Sound and Music Computing*.

[4] J. Janer, J. Bonada, and M. Blaauw, "Performance-driven control for sample-based singing voice synthesis," in *Proceedings of the 9th International Conference on Digital Audio Effects (DAFx)*, Montreal, Canada, 2006.

[5] M. Umbert, J. Bonada, M. Goto, T. Nakano, and J. Sundberg, "Expression control in singing voice synthesis: Features, approaches, evaluation, and challenges," *IEEE Signal Processing Magazine*, vol. 32, no. 6, pp. 55–73, Nov. 2015.

[6] K. Hua. (2018). Modeling singing F0 with neural network driven transition-sustain models. version 1. arXiv: 1803.04030 [eess.AS].

[7] S. Ö. Arik, G. Diamos, A. Gibiansky, J. Miller, K. Peng, W. Ping, J. Raiman, and Y. Zhou, "Deep Voice 2: Multi-speaker neural text-to-speech," in *Advances in Neural Information Processing Systems 30 (NIPS)*, Long Beach, CA, USA, 2017, pp. 2962–2970.

[8] S. Ö. Arik, J. Chen, K. Peng, W. Ping, and Y. Zhou, "Neural voice cloning with a few samples," in *Advances in Neural Information Processing Systems 31 (NIPS)*, Montreal, Canada, 2018.

[9] Y. Chen, Y. Assael, B. Shillingford, D. Budden, S. Reed, H. Zen, Q. Wang, L. C. Cobo, A. Trask, B. Laurie, C. Gulcehre, A. van den Oord, O. Vinyals, and N. de Freitas, "Sample efficient adaptive text-to-speech," in *Proceedings of the 7th International Conference on Learning Representations (ICLR)*, New Orleans, LA, USA, 2019.

[10] M. Morise, F. Yokomori, and K. Ozawa, "WORLD: A vocoder-based high-quality speech synthesis system for real-time applications," *IEICE Transactions on Information and Systems*, vol. E99.D, pp. 1877–1884, 2016.

[11] A. Tamamori, T. Hayashi, K. Kobayashi, K. Takeda, and T. Toda, "Speaker-dependent WaveNet vocoder," in *Proceedings of the 18th Annual Conference of the International Speech Communication Association (Interspeech)*, Stockholm, Sweden, 2017, pp. 1118–1122.

[12] J. Sotelo, S. Mehri, K. Kumar, J. F. Santos, K. Kastner, A. Courville, and Y. Bengio, "Char2Wav: End-to-end speech synthesis," in *Proceedings of the 5th International Conference on Learning Representations (ICLR)*, Toulon, France, 2017.

[13] Y. Jia, Y. Zhang, R. J. Weiss, Q. Wang, J. Shen, F. Ren, Z. Chen, P. Nguyen, R. Pang, I. Lopez-Moreno, and Y. Wu, "Transfer learning from speaker verification to multispeaker text-to-speech synthesis," in *Advances in Neural Information Processing Systems 31 (NIPS)*, Montreal, Canada, 2018.

[14] J. Lorenzo-Trueba, T. Drugman, J. Latorre, T. Merritt, B. Putrycz, and R. Barra-Chicote. (2018). Robust universal neural vocoding. version 1. arXiv: 1811.06292 [eess.AS].

[15] Z. Duan, H. Fang, B. Li, K. C. Sim, and Y. Wang, "The NUS sung and spoken lyrics corpus: A quantitative comparison of singing and speech," in *Asia-Pacific Signal and Information Processing Association Annual Summit and Conference (APSIPA)*, 2013, pp. 1–9.

[16] B. T. Polyak and A. B. Juditsky, "Acceleration of stochastic approximation by averaging," *SIAM Journal on Control and Optimization*, vol. 30, no. 4, pp. 838–855, Jul. 1992.

[17] D. W. Griffin and J. S. Lim, "Signal estimation from modified short-time fourier transform," *IEEE Transactions on Acoustics, Speech, and Signal Processing*, vol. 32, no. 2, pp. 236–243, Apr. 1984.

[18] Y. Taigman, L. Wolf, A. Polyak, and E. Nachmani, "VoiceLoop: Voice fitting and synthesis via a phonological loop," in *Proceedings of the 6th International Conference on Learning Representations (ICLR)*, Vancouver, Canada, 2018.

[19] E. Nachmani, A. Polyak, Y. Taigman, and L. Wolf, "Fitting new speakers based on a short untranscribed sample," in *Proceedings of the 35th International Conference on Machine Learning (ICML)*, Stockholm, Sweden, 2018, pp. 3683–3691.

[20] K. Shirota, K. Nakamura, K. Hashimoto, K. Oura, Y. Nankaku, and K. Tokuda, "Integration of speaker and pitch adaptive training for HMM-based singing voice synthesis," in *Proceedings of the 39th IEEE International Conference on Acoustics, Speech and Signal Processing (ICASSP)*, Florence, Italy, 2014, pp. 2559–2563.